\documentclass[aps,prb,twocolumn]{revtex4}
\usepackage{graphicx}
\usepackage{dcolumn}
\usepackage{amsmath}
\usepackage{amssymb}
\usepackage{epstopdf}
\usepackage{subfigure,amsmath,verbatim,moreverb,bm}
\def\be{\begin{equation}}
\def\ee{\end{equation}}
\def\ber{\begin{eqnarray}}
\def\eer{\end{eqnarray}}

\def\rv{{\bf r}}

\def\fv{{\bf f}}

\newcommand*{\abs}[1]{\lvert#1\rvert}
\newcommand*{\coord}[1]{\mathbf{#1}}

\newcommand*{\du}{\partial}
\DeclareMathOperator{\sgn}{sgn}

\newcommand*{\R}{\mathbb{R}}

\begin{document}

\title{Exchange-correlation functionals from the strongly-interacting limit of DFT:
Applications to model chemical systems}

\author{Francesc Malet, Andr\'e Mirtschink, Klaas J. H. Giesbertz, Lucas O. Wagner, and Paola Gori-Giorgi}
\affiliation{Department of Theoretical Chemistry and
Amsterdam Center for Multiscale Modeling, FEW,Vrije Universiteit,De Boelelaan 1083,1081HV Amsterdam,
The Netherlands}

\begin{abstract}
We study model one-dimensional chemical systems (representative of their three-dimensional counterparts)  using the strictly-correlated electrons (SCE)
functional, which, by construction, becomes asymptotically exact in the limit of infinite coupling strength. The SCE functional has a highly non-local dependence on the density and is able to capture strong correlation within Kohn-Sham theory without introducing any symmetry breaking. 
Chemical systems, however, are not close enough to the strong-interaction limit so that, while ionization energies and the stretched H$_2$ molecule are accurately described, total energies are in general way too low. A correction based on the exact next leading order in the expansion at infinite coupling strength of the Hohenberg-Kohn functional largely improves the results.
\end{abstract}

\maketitle

\section{Introduction}


Despite the enormous success of Kohn-Sham (KS) density functional theory (DFT) \cite{KohSha-PR-65} 
when applied to the study of many chemical systems,\cite{CohMorYan-CR-12} there are still important cases 
for which standard approximate exchange-correlation functionals are inaccurate.\cite{CraTru-PCCP-09,CohMorYan-CR-12,B12}
In particular, systems in which rearrangement of electrons within some (near-degeneracy) or many (strong correlation) partially filled levels is important, such as transition metals, stretched bonds and Mott insulators, represent a big challenge for KS DFT.\cite{CraTru-PCCP-09,CohMorYan-CR-12} 

Similarly to the unrestricted Hartree-Fock method, KS DFT with approximate functionals tries to mimic the physics of strong electronic correlation with symmetry breaking, which, in many cases (but not always), yields reasonable energies.
In complex systems, however, symmetry breaking can occur erratically and can be very sensitive to the functional chosen.\cite{CraTru-PCCP-09}
When many different broken symmetry solutions are competing it becomes difficult to keep the potential energy surfaces continuous. The rigorous KS formulation is also partially lost, and many properties are wrongly characterized.\cite{AniZaaAnd-PRB-91,BorTorKosManAbeRei-IJQC-05}

It is important to keep in mind that Kohn-Sham DFT is, in principle, an exact theory, which should be able to yield the right ground-state density and energy of strongly-correlated systems without resorting to symmetry breaking. The quest for an approximate exchange-correlation functional able to achieve this fundamental goal is a very active research field,\cite{Bec-JCP-13a,Bec-JCP-13b,Bec-JCP-13c,GieLeuBar-PRA-13,CohMorYan-CR-12} which aims at solving what is arguably one of the most important problems in electronic structure theory.

Recently, an alternative approach to the standard way of constructing functionals for KS DFT has been proposed, 
in which the knowledge on the strong-interaction limit of DFT is used to 
build an approximation for the exchange-correlation energy and potential. 
The starting point is the so-called {\em strictly-correlated-electrons} 
(SCE) reference system, introduced by Seidl and 
coworkers,\cite{Sei-PRA-99,GorSeiVig-PRL-09,GorSei-PCCP-10} which has the same 
density as the real interacting one, but in which the electrons are infinitely correlated 
instead of non-interacting. The SCE functional has a highly non-local dependence on the density, but its functional derivative can be easily constructed,\cite{MalGor-PRL-12,MalMirCreReiGor-PRB-13} yielding a local one-body potential which can be used in the
Kohn-Sham scheme to approximate the exchange-correlation term. The SCE functional tends asymptotically to the exact Hartree-exchange-correlation functional in the extreme infinite correlation (or low-density) limit.

Very promisingly, the first applications of this ``KS SCE'' DFT approach, 
performed on one-dimensional (1D) semiconductor quantum wires\cite{MalGor-PRL-12,MalMirCreReiGor-PRB-13} and on two-dimensional quantum dots,\cite{MenMalGor-PRB-14} have shown 
that the SCE exchange-correlation potential is able to describe the physics of 
the strongly-correlated regime within the restricted Kohn-Sham scheme, truly making non-interacting electrons behave as strongly correlated ones. It is thus natural to ask whether with 
this formalism one can also cure the deficiencies of standard DFT approximations in Chemistry.

The physics of strong correlation encoded in the highly non-local density dependence of the SCE functional, however, does not come for free: the SCE problem is sparse but nonlinear, and a general algorithm for its evaluation following the original formulation is still an open problem. Progress has been made recently\cite{MenLin-PRB-13} by using the reformulation of the SCE functional as a mass transportation theory (or optimal transport) problem,\cite{ButDepGor-PRA-12,CotFriKlu-CPAM-13,FriMenPasCotKlu-JCP-13} which allows to evaluate the SCE functional and its functional derivative by means of a maximization under linear constraints, although the procedure is still cumbersome and needs further developments.\cite{MenLin-PRB-13} Probably, it will be necessary to devise approximate ways to deal with the SCE physics in the general three-dimensional case.

Before adventuring into the challenging task of implementing the SCE functional (or approximations thereof) for 
general three-dimensional systems, we feel it is important to understand whether this functional could play a role for Chemistry.
For this reason, we consider here a simple
one-dimensional model that has recently been shown to be a useful laboratory 
to test functionals for chemical problems,\cite{HelFukCasVerMarTokRub-PRA-11,WagStoBurWhi-PCCP-12} offering a reasonably 
close description of the three-dimensional counterparts and being computationally much 
less demanding.

The paper is organized as follows. In Sec.~\ref{s:KSSCE} we briefly describe the KS SCE approach, first at zeroth 
order of approximation and later introducing an higher order correction properly renormalized. In Sec.~\ref{s:model} we present the studied 
one-dimensional models and we describe the details of the actual calculations. The results are then presented and discussed in Sec.~\ref{s:results}. Finally, in Sec.~\ref{s:conclusions} we draw some conclusions and outlook for future works.


\section{The KS SCE approach}\label{s:KSSCE}

As a brief introduction to the KS SCE\cite{MalGor-PRL-12, MalMirCreReiGor-PRB-13} approach, first consider the partitioning of the total energy density functional for an $N$-electron system in the external potential $\hat{V}_{\rm ext}=\sum_{i=1}^Nv_{\rm ext}(\rv_i)$ as $E[\rho] = F[\rho] + \int \rho(\rv)v_{\rm ext}(\rv)d \rv$,  with the internal energy (kinetic plus electron-electron repulsion) expressed as\cite{Lev-PNAS-79, Lie-IJQC-83}
\begin{align}\label{eq_HK}
F[\rho] \equiv \min_{\Psi\to\rho}\langle\Psi|\hat{T} + \hat{V}_{\rm ee}|\Psi\rangle \; ,
\end{align}
where the search is only over ferrmionic wave functions.
It is the delicate interplay of the kinetic energy \emph{and} the electron-electron interaction that makes the evaluation of this functional a daunting task. If the minimization would only contain the kinetic energy,
\begin{align}
T_s[\rho] \equiv \min_{\Psi \to \rho} \langle\Psi|\hat{T}|\Psi\rangle \; ,
\label{eq:Ts}
\end{align}
it is easy to determine its minimum. In particular, the wave function $\Psi$ which achieves this minimum is often a single Slater determinant composed of orbitals that satisfy the Kohn--Sham equations~\cite{KohSha-PR-65}
\begin{equation}
\label{eq_KS}
\left(-\frac{\nabla^2}{2} + v_{\text{ext}}[\rho](\rv) + v_{\text{Hxc}}[\rho](\rv) \right)\phi_i(\rv) 
= \varepsilon_i \phi_i(\rv) \; .
\end{equation}
The Hartree-exchange-correlation potential originates as a Lagrange multiplier to ensure that the Kohn-Sham system yields the required density $\rho$ and is related to the remainder $E_{\text{Hxc}}[\rho] \equiv F[\rho] - T_s[\rho]$ as its functional derivative
\begin{equation}\label{eq_funcderhxc}
\frac{\delta E_{\text{Hxc}}[\rho]}{\delta \rho(\rv)} \equiv v_{\text{Hxc}}[\rho](\rv)\;.
\end{equation}
Seidl and co-workers have demonstrated that the functional $F[\rho]$ can be also explicitly evaluated if it would only contain the electron-electron interaction~\cite{Sei-PRA-99,SeiGorSav-PRA-07,GorVigSei-JCTC-09}
\begin{equation}
\label{eq_VeeSCE}
V_{\rm ee}^{\rm SCE}[\rho]\equiv \min_{\Psi\to\rho}\langle\Psi|\hat{V}_{\rm ee}|\Psi\rangle \;.
\end{equation}
This strictly-correlated-electrons (SCE) functional is the strongly-interacting limit of $F[\rho]$, describing the situation in which the kinetic energy is negligible. The SCE functional is the natural counterpart of the Kohn--Sham kinetic energy as 
it defines, instead of a non-interacting reference system, one in which the electrons 
are infinitely (or perfectly) correlated: if one of the electrons, which can be taken 
as reference and labeled as ``1'', is found with some probability at a given position 
$\rv$, the other $N-1$ electrons will be found, with the same probability, at the 
positions $\rv_i\equiv \fv_i[\rho](\rv)$ $(i=2,...N)$.\cite{SeiGorSav-PRA-07,GorSeiVig-PRL-09,GorSei-PCCP-10} 
Since the $\fv_i$ only depend on $\rv$, they are called \emph{co-motion functions}. They are highly 
non-local functionals of the density and satisfy, for each $\rv$, the set of 
differential equations\cite{SeiGorSav-PRA-07} 
\begin{equation}
\rho(\rv)d \rv = \rho(\fv_i(\rv)) d\fv_i(\rv) \,\,\,\, (i=2,...,N)\; ,
\label{eq_fi}
\end{equation}
as well as the following group properties that ensure that the $N$ electrons are indistinguishable (so that there is no dependence on which electron is chosen as  ``electron 1'')
\begin{equation}
\label{eq_groupprop}
\begin{split}
&\fv_1(\rv) \equiv \rv, \\
&\fv_2(\rv) \equiv \fv(\rv), \\
&\fv_3(\rv) =      \fv(\fv(\rv)), \\
&\fv_4(\rv) =      \fv(\fv(\fv(\rv))), \\
&\qquad\ \vdots \\
&\underbrace{\fv(\fv(\ldots\fv(\fv(\rv))))}_\text{$N$ times} = \rv.
\end{split}
\end{equation}
\begin{figure}[t]
\includegraphics[width=7cm]{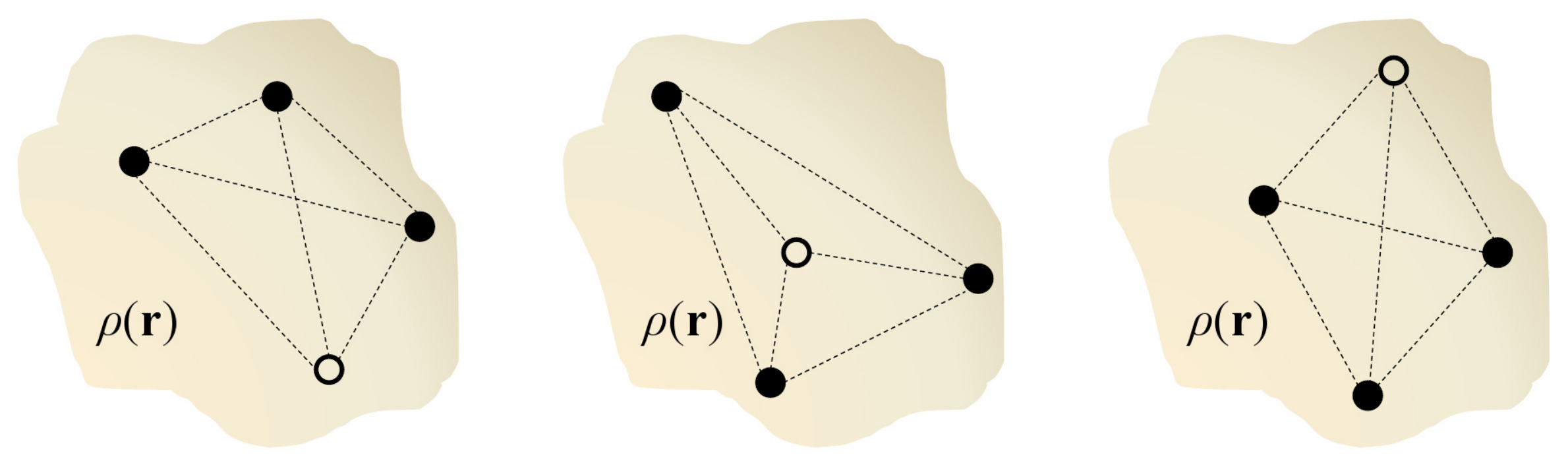}
\caption{Schematic illustration of the SCE reference system for a given density 
$\rho(\mathbf{r})$ and $N=4$ electrons. The empty circle represents the position of 
the reference particle, which is different in each case. The other electrons must 
adapt their relative positions in such a way that the superposition of all the possible 
configurations (one for each $\rv$) yields the density $\rho({\bf r})$.}
\label{fig:schematic}    
\end{figure}%
%
Figure~\ref{fig:schematic} schematically illustrates the SCE reference system for the case of 4 electrons in a 
given density $\rho(\rv)$. The figure shows three configurations, each of them corresponding 
to the reference electron being at a different position $\rv$ (represented by an empty circle). 
The other three electrons adapt their positions, given by the co-motion functions (represented 
by solid symbols), in order to minimize the total Coulomb repulsion and under the constraint 
that the superposition of all the possible configurations (one for each $\rv$) yields 
the density of the physical system $\rho(\rv)$. The SCE system thus represents a smooth 
$N$-electron quantum-mechanical density by means of an infinite superposition of classical 
configurations, which fulfill Eq.~\eqref{eq_fi} for every $\rv$. The square modulus of the 
corresponding SCE wave function (which becomes a distribution in this 
limit\cite{ButDepGor-PRA-12,CotFriKlu-CPAM-13}) can be written as 
\begin{multline}
\label{eq_psi2}
|\Psi_{\rm SCE}(\rv_1,\rv_2,\dots,\rv_N)|^2 = \frac{1}{N!} \sum_{\wp}
\int d\rv \, \frac{\rho(\rv)}{N} \, \delta(\rv_1-\fv_{\wp(1)}(\rv)) \\
{} \times\delta(\rv_2-\fv_{\wp(2)}(\rv)) \cdots \delta(\rv_N-\fv_{\wp(N)}(\rv))\; ,
\end{multline}
where $\wp$ denotes a permutation of ${1,\dots,N}$, such that 
$\rho(\rv) = N \int |\Psi_{\rm SCE}(\rv,\rv_2,\dots,\rv_N)|^2\,d\rv_2\cdots d\rv_N$. The SCE system 
can thus be visualized as a ``floating Wigner crystal'' describing the density $\rho(\rv)$.  

The co-motion functions are the key object for the SCE functional, analogously to the Kohn--Sham orbitals 
for the non-interacting kinetic energy functional $T_s[\rho]$. They can be used to express the functional $V_{\rm ee}^{\rm SCE}[\rho]$ explicitly 
as \cite{SeiGorSav-PRA-07,MirSeiGor-JCTC-12}
\begin{equation}
\label{eq_VeeSCE2}
\begin{split}
V_{\rm ee}^{\rm SCE}[\rho]
&= \int d\rv\,\frac{\rho(\rv)}{N} \sum_{i=1}^{N-1}
\sum_{j=i+1}^N\frac{1}{|\fv_i(\rv)-\fv_j(\rv)|} \\
&= \frac{1}{2}\int d\rv\,\rho(\rv)\sum_{i= 2}^N \frac{1}{|\rv-\fv_i(\rv)|} \; .
\end{split}
\end{equation}
An important property of the SCE system is the following one: since the position of one 
electron at a given $\rv$ determines the other $N-1$ electronic positions, the net Coulomb 
repulsion acting on a certain position $\rv$ becomes a function of $\rv$ itself. As a 
consequence, this force can be written in terms of the negative gradient of some one-body 
local potential $v_{\rm SCE}(\rv)$,\cite{MalMirCreReiGor-PRB-13} such that
\begin{equation}
\label{eq_vSCE}
-{\bf \nabla} v_{\rm SCE}[\rho](\rv)
\equiv {\bf F}_{\rm Coulomb}(\rv)=\sum_{i=2}^N \frac{\rv-\fv_i[\rho](\rv)}{|\rv-\fv_i[\rho](\rv)|^3} \; .
\end{equation}
From the above equation one can see that for a finite system in the limit $|\rv|\rightarrow \infty$ 
the potential $v_{\rm SCE}[\rho](\rv)$ goes as
\begin{equation}
v_{\rm SCE}[\rho](|\rv|\rightarrow\infty) = \frac{N-1}{|\rv|} \; .
\label{eq_vsceasymp}
\end{equation}
Furthermore, it can be shown that it satisfies the important exact 
relation\cite{MalMirCreReiGor-PRB-13}
\begin{equation}
\label{eq_funcder}
v_{\rm SCE}[\rho](\rv) =
\frac{\delta V_{\rm ee}^{\rm SCE}[\rho]}{\delta \rho(\rv)} \; ,
\end{equation}
thus providing a powerful shortcut to the construction of the functional derivative of the SCE functional. 

The zeroth-order ``KS SCE'' approach\cite{MalMirCreReiGor-PRB-13} approximates the Hohenberg-Kohn (HK) functional of Eq.~\eqref{eq_HK} as
\begin{align}
	F[\rho]&\approx \min_{\Psi\rightarrow\rho}\left\langle\Psi | \hat{T}| \Psi \right\rangle + \min_{\Psi\rightarrow\rho}\left\langle\Psi |\hat{V}_{ee}| \Psi \right\rangle\notag\\
	&=T_s[\rho]+V_{ee}^{\rm SCE}[\rho],
	\label{eq:F_KSSCE}
\end{align}
yielding a rigorous lower-bound to the exact energy,\cite{MalGor-PRL-12,MalMirCreReiGor-PRB-13} since
\begin{equation}
T_s[\rho] + V_{\rm ee}^{\rm SCE}[\rho] \leq F[\rho].
\label{eq_Fap}
\end{equation}
Equivalently, comparing Eqs.~\eqref{eq_funcderhxc} and~\eqref{eq_funcder}, we see that the KS SCE
uses  $v_{\text{SCE}}$ to approximate the Hartree and exchange-correlation potentials,
\begin{equation}
v_{\text{Hxc}}(\rv) \simeq v_{\rm SCE}(\rv) \; .
\end{equation}
Notice that Eq.~\eqref{eq_vsceasymp} implies that $v_{\rm SCE}$ as an approximate $v_{\text{Hxc}}$ has the right asymptotics in the limit $|\rv|\rightarrow \infty$. 

Equation~\eqref{eq_Fap} shows that the KS SCE DFT approach treats both the kinetic energy 
and the electron-electron interaction on the same footing, letting the two terms compete in 
a self-consistent way within the Kohn--Sham scheme. Furthermore, the method becomes asymptotically 
exact both in the very weak and very strong correlation 
limits.\cite{MalGor-PRL-12,MalMirCreReiGor-PRB-13} At intermediate 
correlation regimes, however it can seriously underestimate the total energy.\cite{MalMirCreReiGor-PRB-13}

\subsection{Higher-order corrections to zeroth-order KS SCE}
\label{s:ZPEtheory}
In order to discuss corrections to KS SCE, it is useful to rewrite the approximation of Eq.~\eqref{eq:F_KSSCE} in the language of the usual adiabatic connection (coupling-constant integration) of DFT.\cite{LanPer-SSC-75,SeiPerLev-PRA-99}
The HK functional of Eq.~\eqref{eq_HK} and the KS kinetic energy functional of Eq.~\eqref{eq:Ts} can be seen as the value at $\lambda=1$ and $\lambda=0$ of a more general functional $F_\lambda[\rho]$, in which the electronic interaction is rescaled by a coupling strength parameter $\lambda$,
\begin{align}
		F_\lambda[\rho]=\min_{\Psi\rightarrow\rho}\left\langle\Psi | \hat{T} + \lambda\hat{V}_{ee}| \Psi \right\rangle.
		\label{eq:Flambda}
	\end{align}
By denoting $\Psi_{\lambda}[\rho]$ the minimizing wave function in Eq.~\eqref{eq:Flambda}, and by defining
\begin{equation}
	W_{\lambda}[\rho]\equiv \left\langle\Psi_{\lambda}[\rho] |\hat{V}_{ee}| \Psi_{\lambda}[\rho] \right\rangle-E_{\rm Hartree}[\rho],
	\label{eq:Wambdadef}
\end{equation}
one obtains the well-known exact formula\cite{LanPer-SSC-75} for the exchange-correlation functional $E_{xc}[\rho]$,
\begin{equation}
    E_{xc}[\rho]=\int_0^1 W_{\lambda}[\rho]\, d\lambda.
    \label{eq:adiab}
\end{equation}
The functional $V_{ee}^{\rm SCE}[\rho]-E_{\rm Hartree}[\rho]$ is the zeroth-order term in the expansion of $W_\lambda[\rho]$ 
when $\lambda\to\infty$. The next leading term in 
the series is given by\cite{SeiPerKur-PRA-00,GorVigSei-JCTC-09}
\begin{eqnarray}
W_{\lambda\to\infty}[\rho] & = & W_\infty[\rho]+\frac{W'_\infty[\rho]}{\sqrt{\lambda}}+O(\lambda^{-p}) \label{eq_Wbiglambda}\\
W_\infty[\rho] & = & V_{ee}^{\rm SCE}[\rho]-E_{\rm Hartree}[\rho] \\
W'_\infty[\rho] & = & V_{ee}^{\rm ZPE}[\rho],
\label{eq_exp}
\end{eqnarray}
where  ZPE stands for ``zero-point energy'', and $p\geq$5/4 -- see 
Ref.~\cite{GorVigSei-JCTC-09} for further details. 
Physically, the zeroth-order term $V_{ee}^{\rm SCE}[\rho]$ in the expansion~\eqref{eq_exp} corresponds to the 
interaction energy when the electrons are ``frozen'' in the strictly-correlated positions of the SCE floating Wigner 
crystal. The ZPE term in the series takes into account small vibrations of the electrons around their SCE
positions, and it is given by (for electrons in $D$ dimensions)\cite{GorVigSei-JCTC-09}
\begin{equation}
V_{ee}^{\rm ZPE}[\rho]=\frac{1}{2}\int d{\bf r}\frac{\rho({\bf r})}{N}
\sum_{n=1}^{D\,N-D}\frac{\omega_n({\bf r})}{2} \; .
\label{eq_ZPE}
\end{equation}
The $\omega_n({\bf r})$ are the zero-point-energy  vibrational frequencies around the SCE 
minimum,\cite{GorVigSei-JCTC-09} given by the square root of the eigenvalues of the Hessian matrix entering 
the expansion up to second order of the potential energy of the electrons in the SCE system.\cite{GorVigSei-JCTC-09}

\begin{figure}
    \begin{center}
\includegraphics[width=6cm]{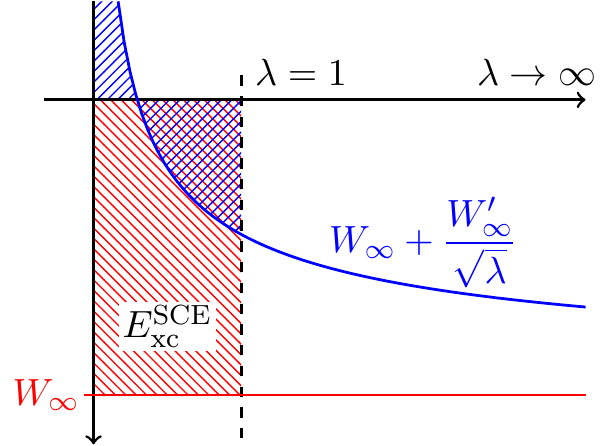}
\includegraphics[width=6cm]{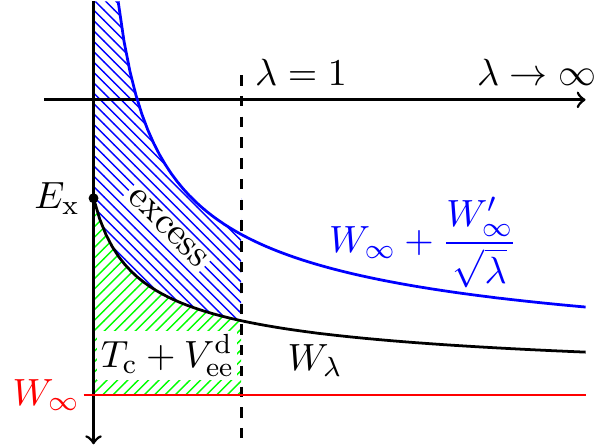}
\end{center}
\caption{Schematic representation of the functional $W_{\lambda}[\rho]$ of Eq.~\eqref{eq:Wambdadef} as a function of $\lambda$. The KS SCE approximation corresponds to setting the exchange-correlation energy equal to the red area in the upper panel. Simply adding the zero point term, corresponds to overcorrecting the KS SCE energy, by including also the  positive blue area in the upper panel. The simple approximation of Eqs.~\eqref{eq:ISIZP1}-\eqref{eq:ISIZPfin} removes the excess energy by shifting the value of $W_{\lambda}[\rho]$ at $\lambda=0$.}
\label{fig:adiab}    
\end{figure}

We see that the KS SCE approximation of Eq.~\eqref{eq:F_KSSCE} corresponds to setting $W_\lambda[\rho]=W_\infty[\rho]$ in the integrand of Eq.~\eqref{eq:adiab}, as schematically shown by the red area in the upper panel of Fig.~\ref{fig:adiab}. 
The exact correction to KS SCE would be the sum of the kinetic correlation energy,  
\begin{equation}
T_c[\rho]=\langle\Psi_{\lambda=1}[\rho]|\hat{T}|\Psi_{\lambda=1}[\rho]\rangle-T_s[\rho],
\end{equation} 
and of the electron-electron decorrelation energy,\cite{LiuBur-JCP-09,GorSeiVig-PRL-09}
\begin{equation}
V_{ee}^d[\rho]=\langle\Psi_{\lambda=1}[\rho]|\hat{V}_{ee}|\Psi_{\lambda=1}[\rho]\rangle-V_{ee}^{\rm SCE}[\rho].
\end{equation} 
If we insert the expansion of Eq.~\eqref{eq_Wbiglambda} into Eq.~\eqref{eq:adiab} we obtain 
\be
T_c[\rho]+V_{ee}^d[\rho]\approx 2\,V_{ee}^{\rm ZPE}[\rho].
\ee
This correction, however, is in general way too large, as it includes the positive contribution to the integrand coming from the integrable divergence $\propto\lambda^{-1/2}$, as represented by the blue area in the upper panel of Fig.~\ref{fig:adiab}.
In order to get a more realistic correction, we consider here a simplified interaction-strength-interpolation (ISI)\cite{SeiPerLev-PRA-99,SeiPerKur-PRL-00} which sets the value of $W_\lambda[\rho]$ at $\lambda=0$ equal to its exact value, the exchange energy $E_x[\rho]$, 
\begin{eqnarray}
	W_{\lambda}^{\rm isiZP}[\rho]& =& W_\infty[\rho]+\frac{W'_\infty[\rho]}{\sqrt{\lambda+a[\rho]}}, \label{eq:ISIZP1}\\
	a[\rho] & = & \left(\frac{W'_\infty[\rho]}{E_x[\rho]-W\infty[\rho]}\right)^2.
\end{eqnarray}
In this way, we remove the excess positive contribution, as shown in the lower panel of Fig.~\ref{fig:adiab}. We then obtain the renormalized correction
\be
T_c[\rho]+V_{ee}^d[\rho]\approx 2\,V_{ee}^{\rm ZPE}[\rho]\left(\sqrt{1+a[\rho]}-\sqrt{a[\rho]}\right).
\label{eq:ISIZPfin}
\ee
This correction is size consistent only when a system dissociates into equal fragments. A full size-consistent approximation would require a {\em local} interpolation along the adiabatic connection, as discussed in Ref.~\cite{MirSeiGor-JCTC-12}.

\section{Details of the calculations for one-dimensional model systems}\label{s:model}

We have considered one-dimensional (1D) models for different atoms and ions and for the H$_2$ molecule. 
In order to avoid the divergence of the Coulomb interaction in one dimension, we consider the 
electrons and the nuclei to interact via a soft-Coulomb potential given by
\begin{equation}
v_{\rm soft}(x)=\frac{q_1q_2}{\sqrt{1 + x^2}} \; ,
\label{eq_softc}
\end{equation}
where $q_1$ and $q_2$ are the charges of the nuclei/electrons. For the calculations presented here we 
have considered nuclear charges $q_1=$1, 2, 3 and 4, corresponding to the elements H, He, Li and Be, 
respectively. In each case we have also studied different ionic species. Finally, for the H$_2$ molecule, 
we have considered interatomic separations $R_{\rm H-H}$ in the range between 0 and 20 atomic units.
\begin{figure}
    \begin{center}
\includegraphics[width=7cm]{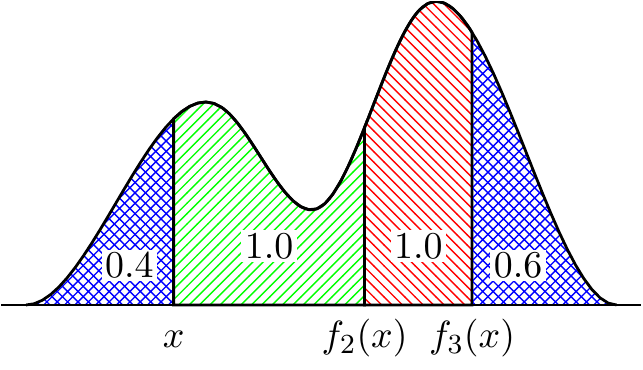}
\end{center}
\caption{Schematic illustration of the co-motion functions in 1D. Two adjacent strictly correlated positions are always separated by a distance such that the density between them integrates exactly to one electron.}
\label{fig:f1D}    
\end{figure}
\subsection{Calculation of the co-motion functions and of the SCE potential}
In one dimension, the co-motion functions $f_i(x)$ can be calculated analytically by integrating 
Eqs.~\eqref{eq_fi} for a given density $\rho(x)$,\cite{Sei-PRA-99,RasSeiGor-PRB-11,ButDepGor-PRA-12} 
choosing boundary conditions that make the density between two adjacent strictly-correlated positions 
always integrate to 1 (total suppression of fluctuations),\cite{Sei-PRA-99} as schematically illustrated in Fig.~\ref{fig:f1D},
\be
\int_{f_i(x)}^{f_{i+1}(x)}\rho(x')\,dx'=1,
\ee
and ensuring that the $f_i(x)$ satisfy the required group 
properties of Eq.~\eqref{eq_groupprop}.\cite{Sei-PRA-99,SeiGorSav-PRA-07,ButDepGor-PRA-12} This yields
\be
f_i(x)=
\begin{cases}
N_e^{-1}[N_e(x)+i-1]		& x \leq a_{N+1-i} \\
N_e^{-1}[N_e(x)+i-1-N]	& x > a_{N+1-i},
\end{cases}
\label{eq:f1Dformulas}
\ee
where the function $N_e(x)$ is defined as
\be
N_e(x)=\int_{-\infty}^x\rho(x')\,dx',
\label{eq:Ne}
\ee
and $a_{k}=N_e^{-1}(k)$. 
Equation \eqref{eq_vSCE} becomes in this case
\be
v'_{\rm SCE}[\rho](x)=\sum_{i=2}^N w'(|x-f_i(x)|)\sgn(x-f_i(x)),
\label{eq_vSCE1D}
\ee
where $w(x)$ denotes the interaction between the particles, which will be the soft-Coulomb interaction~\eqref{eq_softc} in our case, and $w'(x)$ is its derivative. Notice the highly non-local dependence of the co-motion functions on $\rho(x)$, as clearly shown by Eqs.~\eqref{eq:f1Dformulas}-\eqref{eq:Ne}, and the great simplification of the functional derivative of the SCE functional provided by Eq.~\eqref{eq_vSCE1D}.

To perform practical calculations, one must proceed self-consistently in three 
steps: i) generate the co-motion functions via Eqs.~\eqref{eq:f1Dformulas}-\eqref{eq:Ne} for a given density $\rho(x)$; ii) calculate $v_{\rm SCE}(x)$ by integrating Eq.~\eqref{eq_vSCE1D} 
with the boundary condition $v_{\rm SCE}[\rho](|x|\to\infty)=0$;  iii) use 
the approximation $v_{\rm Hxc}(x)\approx v_{\rm SCE}(x)$ to solve the Kohn-Sham equations~\eqref{eq_KS}. The total energy is then obtained by adding the external potential contribution to Eq.~\eqref{eq:F_KSSCE}. As said, we work in the original spin-restricted KS framework, in which 
each spatial orbital is doubly occupied.

\subsection{Calculation of the Zero-Point Energies (ZPE) in 1D}

To compute the ZPE we start from the classical energy expression of the SCE system as a function of the coordinates of the individual electrons. Since the kinetic energy in the $\lambda \to \infty$ limit becomes infinitely small, only the potential parts remain: the interaction between the electrons and the SCE potential
\be
E_{\text{pot}}(x_1,\dotsc,x_N) = \sum_{i=1}^{N-1}\sum_{j=i+1}^Nw(\abs{x_i-x_j}) - \sum_{i=1}^Nv_{\text{SCE}}(x_i).
\ee
Notice that the SCE potential counteracts exactly the repulsive forces due to the interaction, making $E_{\text{pot}}$ stationary on the 1D subspace of $\R^N$ parametrized by $\coord{f}(x)\equiv \{x_1 = f_1(x)=x, x_2 = f_2(x), \dotsc, x_N = f_N(x)\}$. The diagonal contributions to the Hessian are readily evaluated to be
\be
\du_{x_i}^2E_{\text{pot}}\bigl(\coord{f}(x)\bigr)
= \sum_{k \neq i}^Nw''(\abs{f_i(x) - f_k(x)})\frac{\rho\bigl(f_i(x)\bigr)}{\rho\bigl(f_k(x)\bigr)}
\ee
and the off-diagonal elements become ($i \neq j$)
\be
\du_{x_i}\du_{x_j}E_{\text{pot}}\bigl(\coord{f}(x)\bigr) = -w''(\abs{f_i(x) - f_j(x)}).
\ee
In the special case of a two-electron system, the matrix elements of the Hessian simplify to
\begin{align}
\du_{x_1}^2E_{\text{pot}}\bigl(x,f(x)\bigr)			&= w''(\abs{x - f(x)})\frac{\rho(x)}{\rho\bigl(f(x)\bigr)}, \notag \\
\du_{x_2}^2E_{\text{pot}}\bigl(x,f(x)\bigr)			&= w''(\abs{x - f(x)})\frac{\rho\bigl(f(x)\bigr)}{\rho(x)}, \\
\du_{x_1}\du_{x_2}E_{\text{pot}}\bigl(x,f(x)\bigr)		&= -w''(\abs{x - f(x)}), \notag
\end{align}
where we used $f_1(x) = x$ and have set $f(x) = f_2(x)$. The Hessian is readily diagonalized and gives the zero-point frequency
\begin{align}
\omega(x) = \sqrt{w''\bigl(\abs{x - f(x)}\bigr)\left(\frac{\rho(x)}{\rho\bigl(f(x)\bigr)} + \frac{\rho\bigl(f(x)\bigr)}{\rho(x)}\right)},
\label{eq:wN2}
\end{align}
which can be used in the ZPE expression~\eqref{eq_ZPE} to calculate the ZPE correction.
The other eigenvalue is zero as expected, since the SCE system corresponds to a floating Wigner crystal, and $E_{\text{pot}}$ should be degenerate on $\{\coord{f}(x) \mid x \in \R\}$. Hence, the energy surface is flat in this direction which gives a vanishing eigenvalue in the Hessian of the classical potential energy.

For $N>2$ we have diagonalized numerically the Hessian matrix on our grid. This needs to be done just in one interval between two adjacent $a_i=N_e^{-1}(i)$, e.g. for $x\in[a_1,a_2]$, because the properties of the co-motion functions ensure that 
\begin{equation}
\int_{-\infty}^\infty dx\frac{\rho(x)}{N}
\sum_{n=1}^{N-1} \omega_n(x)= \int_{a_i}^{a_{i+1}} dx\, \rho(x)
\sum_{n=1}^{N-1}\omega_n(x).
\label{eq_ZPE1D}
\end{equation}

A warning should be added because the soft Coulomb interaction $w(x)=v_{\rm soft}(x)$ of Eq.~\eqref{eq_softc} is not convex for $x<2^{-1/2}$. As it is evident from the formulas for $N=2$ of Eq.~\eqref{eq:wN2}, the ZPE breaks down when $w''(|x-f(x)|)<0$. For all the systems studied, the minimum possible SCE electron-electron distance is always larger than $2^{-1/2}$, so that the nonconvexity of $w(x)$ does not pose any problem.

\subsection{Comparison with other approaches}

We validate our results by comparing them with those obtained from the density matrix renormalization group (DMRG) as described in Ref.~\cite{WagStoBurWhi-PCCP-12}. For the H$_2$ potential energy curve, we have also carried out full CI calculations on a numerical grid. We have also performed, for comparison, calculations with 
the Kohn-Sham local-density-approximation in both the spin-restricted (LDA)
and spin-unrestricted (LSDA) formulations. The parametrization of the
L(S)DA exchange-correlation functional with soft-coulomb interaction is taken from Ref.~\cite{HelFukCasVerMarTokRub-PRA-11}.


\section{Results}
\label{s:results}

\subsection{1D atoms and ions}

Table~\ref{table:energies} shows the total energies for different atomic elements, comparing the results
obtained with the KS SCE, DMRG and KS L(S)DA approaches. The renormalized ZPE correction ``isiZP'' of Eq.~\eqref{eq:ISIZPfin} has been added at the postfunctional level to the KS SCE self-consistent energies.
We see that KS SCE largely underestimates all the total energies (except for the $N=1$ systems for which it is exact), providing a lower bound that it is not very tight, and resulting in an accuracy worse than the one of L(S)DA. The isiZP correction improves the results consistently, getting much closer to the DMRG calculations than L(S)DA. 

Particularly interesting is the case of the negative ions, that are a notorious problem in approximate KS DFT. Similarly to the 3D case, the anions are all not bound in L(S)DA, while KS SCE overbinds, yielding a bound system also for He$^{-}$ and Li$^{-}$, which are unbound in DMRG. The isiZP correction, however, correctly predicts H$^-$ to be bound (with a rather good energy) and He$^{-}$ and Li$^{-}$  to be unbound (as their energy becomes higher than the one of the corresponding neutral system). Notice that in Ref.~\cite{HelFukCasVerMarTokRub-PRA-11} the 1D He$^{-}$ and Li$^{-}$ were predicted to be bound, although the value reported for the energy of Li$^-$ was higher than the one of Li. With DMRG we found these two 1D anions to be unbound.

\begin{table}[t]
\small
  \caption{Total energies obtained with the different approaches.
           The DMRG data are from Ref.~\cite{WagStoBurWhi-PCCP-12}}
  \label{table:energies}                                         
\begin{tabular*}{\columnwidth}{@{\extracolsep{\fill}}lccccc}
    \hline
    & KS SCE & DMRG & LDA & LSDA & KS SCE + isiZPE \\
    \hline
    H        & -0.67 & -0.67 & -0.60 & -0.65 & -0.67\\
    H$^{-}$   & -0.89 & -0.73 & - & - & -0.75\\
    He       & -2.38  & -2.24 & -2.20 & -2.20 & -2.25\\
    He$^{-}$  & -2.42 & - & - & - & 	-2.22\\
    He$^{+}$  & -1.48 & -1.48 & -1.41 & -1.45 & -1.48\\
    Li       & -4.43 & -4.21 & -4.16 & -4.18 & -4.22\\ 
    Li$^{-}$  & -4.51 & - & - & - & -4.19\\
    Li$^{+}$  & -4.02 & -3.90 & -3.85 & -3.85 & -3.92\\
    Li$^{2+}$  & -2.34 & -2.34 & -2.26 & -2.30 & -2.34\\
    Be        & -7.12 & -6.79 & -6.76 & -6.76 & -6.79\\ 
    Be$^{+}$   & -6.65 & -6.45 & -6.39 & -6.41 & -6.47\\
    Be$^{2+}$  & -5.72 & -5.62 & -5.56 & -5.56 & -5.64\\
    Be$^{3+}$  & -3.21 & -3.21 & -3.13 & -3.18 & -3.21\\ 
    \hline
  \end{tabular*}
\end{table}

Table~\ref{table:ionization} is the analogue of Table~\ref{table:energies} comparing the negative of the highest occupied (HOMO) KS eigenvalues with the ionization energies obtained from DMRG. We see that the KS SCE HOMO yields quite accurate estimates of the ionization energies, thanks to the right asymptotic behavior of the SCE potential. 

\begin{table}[t]
\small
  \caption{ Same as Table~\ref{table:energies} for the ionization energies (using the
   HOMO eigenvalue for the DFT approaches)}
  \label{table:ionization}
  \begin{tabular*}{\columnwidth}{@{\extracolsep{\fill}}lccccc}
    \hline
    & $-\varepsilon_{\rm HOMO}^{\rm KS SCE}$ & DMRG & $-\varepsilon_{\rm HOMO}^{\rm LDA}$ & $-\varepsilon_{\rm HOMO}^{\rm LSDA}$\\
    \hline
    H        & 0.67  & 0.67 & 0.35 & 0.41 \\
    H$^{-}$   & 0.089 & 0.06 & - &  - \\
    He       & 0.72 & 0.75 & 0.48 & 0.48 \\
    He$^{+}$  & 1.48 & 1.48 & 1.12 & 1.18 \\
    Li       & 0.32 & 0.31 & 0.14 & 0.17 \\ 
    Li$^{+}$  & 1.50 & 1.56 & 1.24 & 1.24 \\
    Li$^{2+}$ & 2.34 & 2.34 & 1.95 & 2.00 \\
    Be       & 0.34 & 0.34 & 0.16 & 0.16 \\ 
    Be$^{+}$  & 0.81 & 0.83 & 0.60 & 0.63 \\
    Be$^{2+}$ & 2.34 & 2.41 & 2.06 & 2.06 \\
    Be$^{3+}$ & 3.21 & 3.21 & 2.81 & 2.86 \\ 
    \hline
  \end{tabular*}
\end{table}

\subsection{1D H$_2$ molecule}

\begin{figure}[t]
\centering
  \includegraphics[width=7cm]{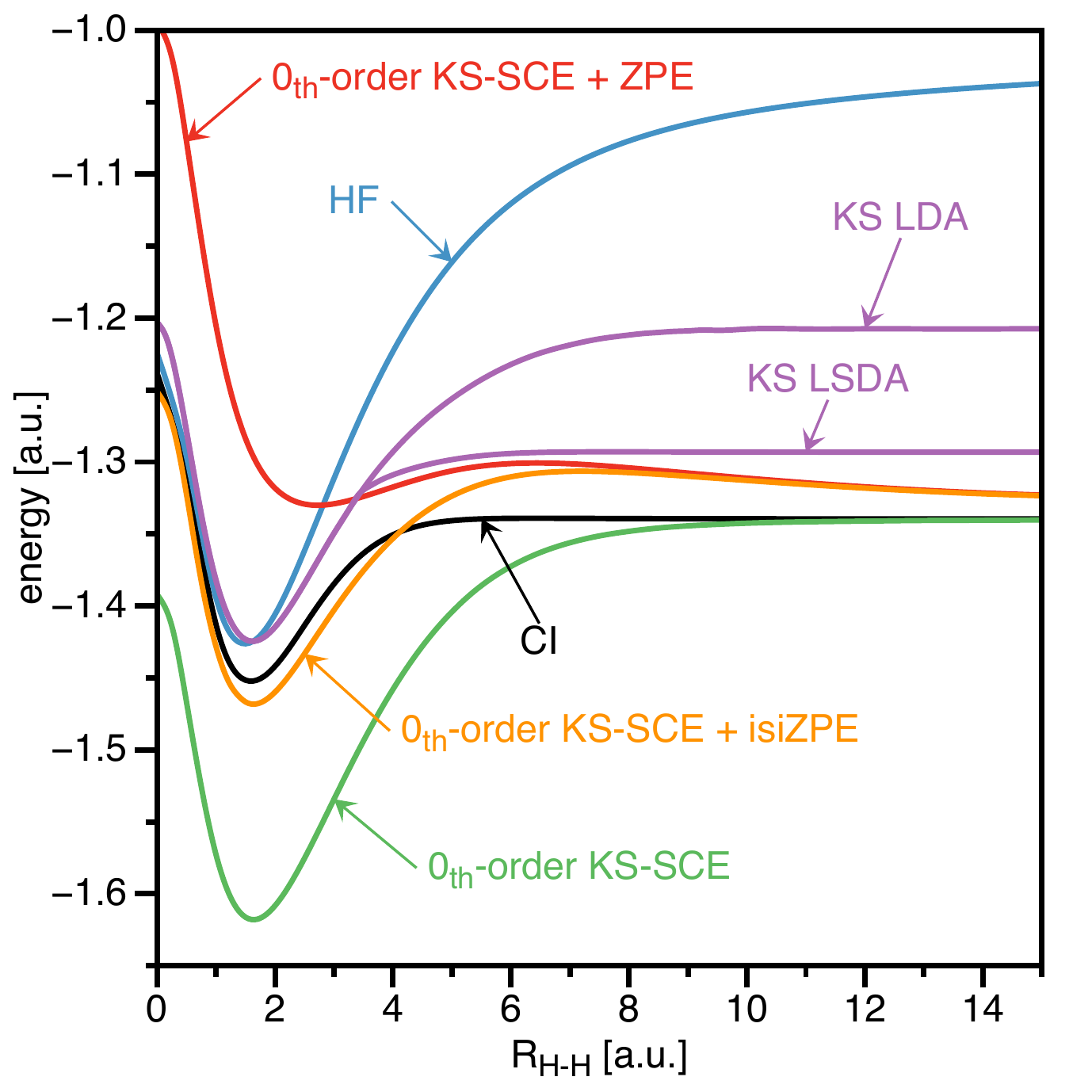}
  \caption{Dissociation energy curves for H$_2$ corresponding to the different
   approaches.}
  \label{fig_H2}
\end{figure}

Figure~\ref{fig_H2} shows the dissociation energy curves obtained from the various methods.  One can see that 
whereas the KS LDA, KS LSDA and restricted Hartree-Fock (HF) energies are relatively close to the CI values near equilibrium, the KS SCE approach
yields a large error due to its overestimation of the electronic correlation.
As the interatomic distance increases, however, one can see that while the spin-resticted 
LDA and HF energies as usual become too positive,
the KS SCE result becomes now increasingly more accurate, tending to the exact curve 
in the dissociation limit. The ability to correctly describe
this limit is remarkable in a spin-restricted 
formalism. 

In the figure we also show the energy curve obtained when the full ZPE correction and the renormalized isiZP
corrections are added, at a postfunctional level, to the zeroth-order KS SCE energies.
One can see that, as expected from the discussion in Sec.~\ref{s:ZPEtheory}, the addition of the bare ZPE gives energies way too high. The inclusion of the isiZP, instead, gives very good results for $R_{\rm H-H}\lesssim 4$~a.u, but displays a ``bump'' in the potential energy curve, which now tends from above to the exact dissociation limit, reached only at $R_{\rm H-H}\gtrsim 20$.

\begin{figure}[t]
\centering
  \includegraphics[width=7cm]{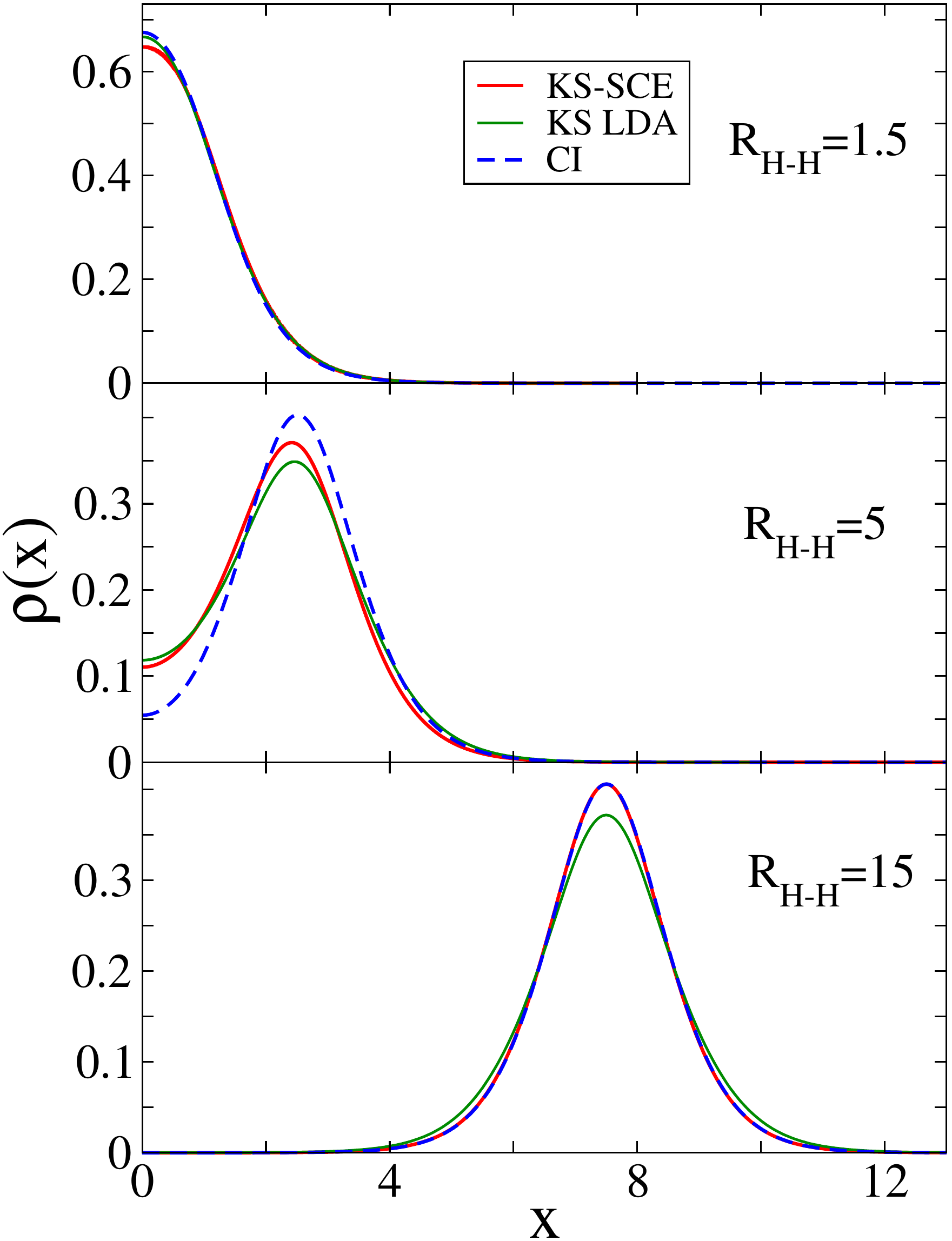}
  \caption{Densities for the H$_2$ molecule corresponding to different
   interatomic separations obtained with the KS SCE, LDA, and CI
   approaches.}
  \label{fig:densH2}
\end{figure}

Figure~\ref{fig:densH2} shows the electronic densities obtained with the KS SCE, KS LDA,
and CI approaches for different interatomic separations $R_{\rm H-H}$. One can see that
for $R_{\rm H-H}=1.5$, which  corresponds to a near-equilibrium
configuration, the KS SCE density is slightly less peaked at the midbond due to 
the above-mentioned overestimation of the correlation. The LDA approach
shows a very good agreement with the exact result. As the interatomic 
separation increases, here with $R_{\rm H-H}=5$, the LDA largely overestimates the density at the
midpoint, clearly reflecting its inability to properly describe the molecular 
dissociation process. The KS SCE approach, instead, shows an improving tendency in
the direction of the exact result, in accordance with the energy curve of Fig.~\ref{fig_H2}.
Finally, in the dissociation limit, represented in the figure by $R_{\rm H-H}=15$, the agreement 
between the KS SCE and CI densities is excellent, whereas the spin-restricted LDA
density is too delocalized.

\begin{figure}[t]
\begin{center}
  \includegraphics[width=7cm]{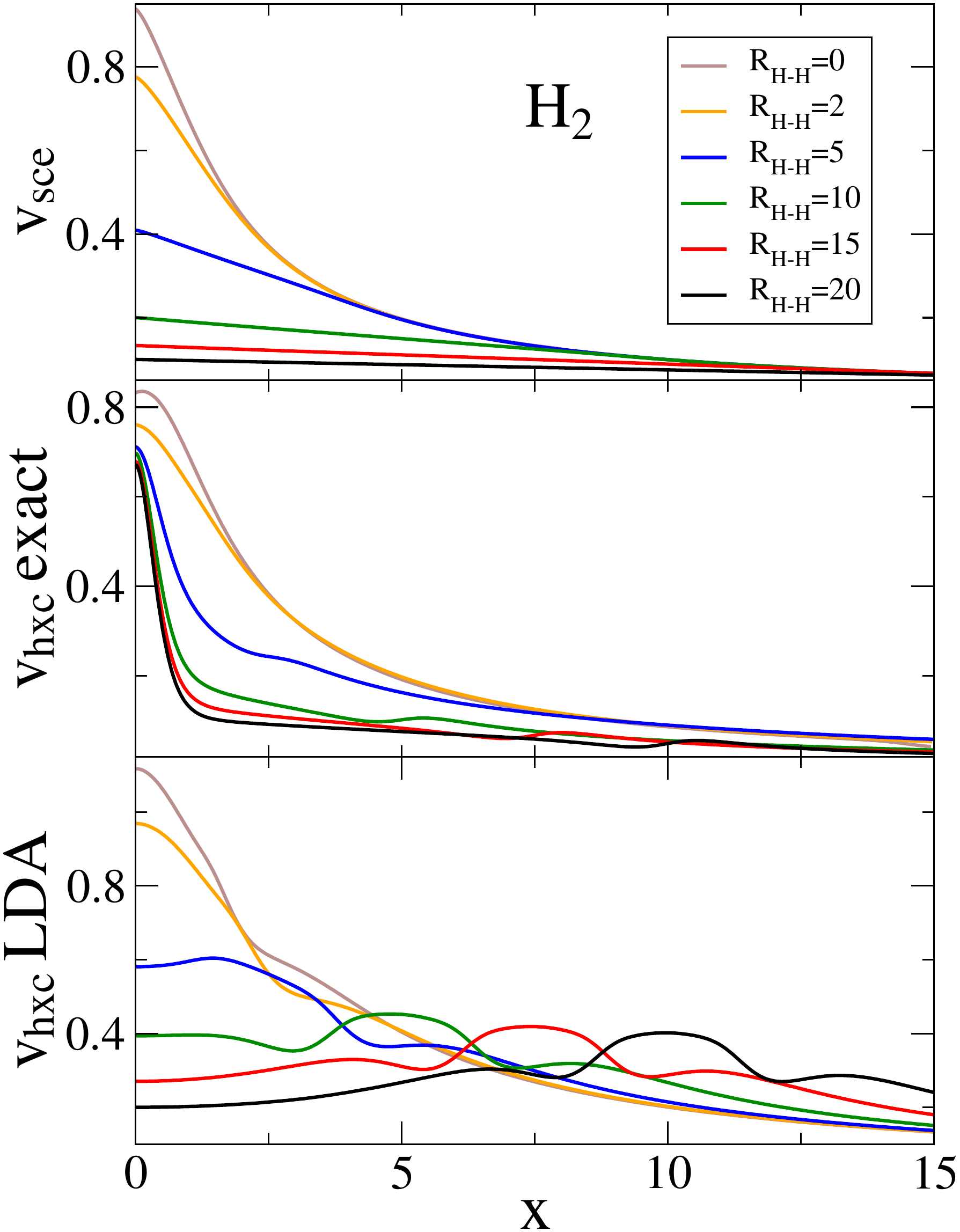}
  \end{center}
  \caption{Exact Kohn-Sham potential (by inversion), Hatree-exchange-correlation 
   potential from KS LDA, and $v_{\rm SCE}$ potential for different interatomic 
   separations $R_{\rm H-H}$ for the 1D H$_2$ molecule.}
  \label{fig:bump}
\end{figure}

The rapid decrease of the exact density at the midbond as the interatomic separation $R_{\rm H-H}$ 
increases is related to the existence of a barrier in the 
corresponding Kohn-Sham potential,\cite{BuiBaeSni-PRA-89,HelTokRub-JCP-09} which 
we show in Fig.~\ref{fig:bump}. The exact barrier has a component that is known to saturate for large internuclear distances $R_{\rm H-H}$ with a height determined by the ionization potential.\cite{HelTokRub-JCP-09} This component is due to the kinetic correlation energy\cite{BuiBaeSni-PRA-89,HelTokRub-JCP-09}  and is thus not captured by the SCE functional that lacks the kinetic energy contribution. The KS SCE barrier, thus, decreases when $R_{\rm H-H}$ increases, and becomes small at large $R_{\rm H-H}$. However, at large internuclear distances, the energetic contribution of the barrier is negligible so that even a  very small barrier (as the one obtained in KS SCE) is enough to get an accurate localized density and the correct energy at dissociation. 
When the confining potential is harmonic, as in the quantum wires and quantum dots studied in Ref.~\cite{MalGor-PRL-12,MalMirCreReiGor-PRB-13,MenMalGor-PRB-14}, the barriers remain finite in the KS SCE potential also at very low density.
In LDA we see that at large $R_{\rm H-H}$ there is a barrier localized on the atoms rather than in the midbond, leading to overestimation of the charge in the bond.

\section{Conclusions and Perspectives}\label{s:conclusions}
The strictly-correlated electrons functional provides an alternative route to the construction of approximations for KS DFT. Instead of following the standard path of including more and more ingredients (``Jacob's ladder'')\cite{PerSmi-INC-01,PerRuzTaoStaScuCso-JCP-05} such as  the local density, the local density gradients, the Kohn-Sham local kinetic energies, etc., the new ingredient here is the non-locality encoded in the SCE functional and higher-order corrections.

While for low-dimensional nanodevices, which can reach very low densities because of their stronger confining potentials, the KS SCE approach is very accurate,\cite{MalMirCreReiGor-PRB-13,MenMalGor-PRB-14} chemical systems are in general not close enough to the strong-interaction limit, so that total energies are too low. We see, however, that the SCE functional is able to capture the strong correlation of a stretched bond, and, with a correction renormalized with exact exchange, to yield accurate results for total energies, predicting the delicate physics of negative ions.
Thanks to the right asymptotic properties of the SCE potential, the KS SCE HOMO also yields an accurate estimate of ionization energies.

Overall, it seems promising to use the SCE physics as an ingredient to build approximate functionals. An exact evaluation of the co-motion functions in the general 3D case might turn out to be too demanding (although progress has been done recently through a different approach\cite{MenLin-PRB-13}). However, it should be possible to build {\em approximate} co-motion functions or, more generally, non-local functionals inspired to the SCE mathematical structure. 

It seems also clear that the accuracy of SCE is somewhat complementary to the ones of standard functionals, so that corrections to SCE which either include exact exchange (like the simple one presented here), or based on standard approximations can be pursued in the future. In general, given an approximate exchange-correlation functional $E_{xc}^{\rm approx}[\rho]$ it is possible to extract from it a correction to KS SCE by using the scaling 
properties \cite{LevPer-PRA-85,PerTaoStaScu-JCP-04} of DFT. By defining, for electrons 
in $D$ dimensions, 
a scaled density $\rho_\gamma(\rv)\equiv\gamma^D\rho(\gamma\, \rv)$ with $\gamma > 0$, we 
have \cite{PerTaoStaScu-JCP-04}
\be
T_c[\rho]+V_{ee}^d[\rho]\approx E_{xc}^{\rm approx}[\rho]-\lim_{\gamma\to 0} \frac{1}{\gamma}E_{xc}^{\rm approx}[\rho_\gamma].
\ee
This way of constructing corrections to KS SCE has been tested by using the LDA functional in 
Refs.~\cite{MalMirCreReiGor-PRB-13} for quantum wires and very recently in Ref.~\cite{MirUmrMorGor-JCP-14} for the anions of the He isoloectronic series. While in the former case they gave very disappointing results, in the latter they improved the results considerably, showing that for chemical problems this could be a good way to proceed. 

\section*{Acknowledgements}
P.G-G. and K.J.H.G. acknowledge support from the Netherlands Organization for Scientific 
Research (NWO) through a VIDI and a VENI grant, respectively. This research was also 
supported by a Marie Curie Intra European Fellowship within the 7th European Community 
Framework Programme (F.M.).

\providecommand*{\mcitethebibliography}{\thebibliography}
\csname @ifundefined\endcsname{endmcitethebibliography}
{\let\endmcitethebibliography\endthebibliography}{}


\begin{mcitethebibliography}{40}
\providecommand*{\natexlab}[1]{#1}
\providecommand*{\mciteSetBstSublistMode}[1]{}
\providecommand*{\mciteSetBstMaxWidthForm}[2]{}
\providecommand*{\mciteBstWouldAddEndPuncttrue}
  {\def\EndOfBibitem{\unskip.}}
\providecommand*{\mciteBstWouldAddEndPunctfalse}
  {\let\EndOfBibitem\relax}
\providecommand*{\mciteSetBstMidEndSepPunct}[3]{}
\providecommand*{\mciteSetBstSublistLabelBeginEnd}[3]{}
\providecommand*{\EndOfBibitem}{}
\mciteSetBstSublistMode{f}
\mciteSetBstMaxWidthForm{subitem}
{(\emph{\alph{mcitesubitemcount}})}
\mciteSetBstSublistLabelBeginEnd{\mcitemaxwidthsubitemform\space}
{\relax}{\relax}

\bibitem[Kohn and Sham(1965)]{KohSha-PR-65}
W.~Kohn and L.~J. Sham, \emph{Phys. Rev. A}, 1965, \textbf{140}, 1133\relax
\mciteBstWouldAddEndPuncttrue
\mciteSetBstMidEndSepPunct{\mcitedefaultmidpunct}
{\mcitedefaultendpunct}{\mcitedefaultseppunct}\relax
\EndOfBibitem
\bibitem[Cohen \emph{et~al.}(2012)Cohen, Mori-S\'anchez, and
  Yang]{CohMorYan-CR-12}
A.~J. Cohen, P.~Mori-S\'anchez and W.~Yang, \emph{Chem. Rev.}, 2012,
  \textbf{112}, 289\relax
\mciteBstWouldAddEndPuncttrue
\mciteSetBstMidEndSepPunct{\mcitedefaultmidpunct}
{\mcitedefaultendpunct}{\mcitedefaultseppunct}\relax
\EndOfBibitem
\bibitem[Cramer and Truhlar(2009)]{CraTru-PCCP-09}
C.~J. Cramer and D.~G. Truhlar, \emph{Phys. Chem. Chem. Phys.}, 2009,
  \textbf{11}, 10757\relax
\mciteBstWouldAddEndPuncttrue
\mciteSetBstMidEndSepPunct{\mcitedefaultmidpunct}
{\mcitedefaultendpunct}{\mcitedefaultseppunct}\relax
\EndOfBibitem
\bibitem[Burke(2012)]{B12}
K.~Burke, \emph{J. Chem. Phys.}, 2012, \textbf{136}, 150901\relax
\mciteBstWouldAddEndPuncttrue
\mciteSetBstMidEndSepPunct{\mcitedefaultmidpunct}
{\mcitedefaultendpunct}{\mcitedefaultseppunct}\relax
\EndOfBibitem
\bibitem[Anisimov \emph{et~al.}(1991)Anisimov, Zaanen, and
  Andersen]{AniZaaAnd-PRB-91}
V.~I. Anisimov, J.~Zaanen and O.~K. Andersen, \emph{Phys. Rev. B}, 1991,
  \textbf{44}, 943\relax
\mciteBstWouldAddEndPuncttrue
\mciteSetBstMidEndSepPunct{\mcitedefaultmidpunct}
{\mcitedefaultendpunct}{\mcitedefaultseppunct}\relax
\EndOfBibitem
\bibitem[Borgh \emph{et~al.}(2005)Borgh, Toreblad, Koskinen, Manninen, Aberg,
  and Reimann]{BorTorKosManAbeRei-IJQC-05}
M.~Borgh, M.~Toreblad, M.~Koskinen, M.~Manninen, S.~Aberg and S.~M. Reimann,
  \emph{Int. J. Quantum Chem.}, 2005, \textbf{{105}}, 817\relax
\mciteBstWouldAddEndPuncttrue
\mciteSetBstMidEndSepPunct{\mcitedefaultmidpunct}
{\mcitedefaultendpunct}{\mcitedefaultseppunct}\relax
\EndOfBibitem
\bibitem[Becke(2013)]{Bec-JCP-13a}
A.~D. Becke, \emph{J. Chem. Phys.}, 2013, \textbf{{138}}, 074109\relax
\mciteBstWouldAddEndPuncttrue
\mciteSetBstMidEndSepPunct{\mcitedefaultmidpunct}
{\mcitedefaultendpunct}{\mcitedefaultseppunct}\relax
\EndOfBibitem
\bibitem[Becke(2013)]{Bec-JCP-13b}
A.~D. Becke, \emph{J. Chem. Phys.}, 2013, \textbf{{138}}, 161101\relax
\mciteBstWouldAddEndPuncttrue
\mciteSetBstMidEndSepPunct{\mcitedefaultmidpunct}
{\mcitedefaultendpunct}{\mcitedefaultseppunct}\relax
\EndOfBibitem
\bibitem[Becke(2013)]{Bec-JCP-13c}
A.~D. Becke, \emph{J. Chem. Phys.}, 2013, \textbf{{139}}, 021104\relax
\mciteBstWouldAddEndPuncttrue
\mciteSetBstMidEndSepPunct{\mcitedefaultmidpunct}
{\mcitedefaultendpunct}{\mcitedefaultseppunct}\relax
\EndOfBibitem
\bibitem[Giesbertz \emph{et~al.}(2013)Giesbertz, {van Leeuwen}, and {von
  Barth}]{GieLeuBar-PRA-13}
K.~J.~H. Giesbertz, R.~{van Leeuwen} and U.~{von Barth}, \emph{Phys. Rev. A},
  2013, \textbf{87}, 022514\relax
\mciteBstWouldAddEndPuncttrue
\mciteSetBstMidEndSepPunct{\mcitedefaultmidpunct}
{\mcitedefaultendpunct}{\mcitedefaultseppunct}\relax
\EndOfBibitem
\bibitem[Seidl(1999)]{Sei-PRA-99}
M.~Seidl, \emph{Phys. Rev. A}, 1999, \textbf{{60}}, 4387\relax
\mciteBstWouldAddEndPuncttrue
\mciteSetBstMidEndSepPunct{\mcitedefaultmidpunct}
{\mcitedefaultendpunct}{\mcitedefaultseppunct}\relax
\EndOfBibitem
\bibitem[Gori-Giorgi \emph{et~al.}(2009)Gori-Giorgi, Seidl, and
  Vignale]{GorSeiVig-PRL-09}
P.~Gori-Giorgi, M.~Seidl and G.~Vignale, \emph{Phys. Rev. Lett.}, 2009,
  \textbf{{103}}, 166402\relax
\mciteBstWouldAddEndPuncttrue
\mciteSetBstMidEndSepPunct{\mcitedefaultmidpunct}
{\mcitedefaultendpunct}{\mcitedefaultseppunct}\relax
\EndOfBibitem
\bibitem[Gori-Giorgi and Seidl(2010)]{GorSei-PCCP-10}
P.~Gori-Giorgi and M.~Seidl, \emph{Phys. Chem. Chem. Phys.}, 2010,
  \textbf{{12}}, 14405\relax
\mciteBstWouldAddEndPuncttrue
\mciteSetBstMidEndSepPunct{\mcitedefaultmidpunct}
{\mcitedefaultendpunct}{\mcitedefaultseppunct}\relax
\EndOfBibitem
\bibitem[Malet and Gori-Giorgi(2012)]{MalGor-PRL-12}
F.~Malet and P.~Gori-Giorgi, \emph{Phys. Rev. Lett.}, 2012, \textbf{{109}},
  246402\relax
\mciteBstWouldAddEndPuncttrue
\mciteSetBstMidEndSepPunct{\mcitedefaultmidpunct}
{\mcitedefaultendpunct}{\mcitedefaultseppunct}\relax
\EndOfBibitem
\bibitem[Malet \emph{et~al.}(2013)Malet, Mirtschink, Cremon, Reimann, and
  Gori-Giorgi]{MalMirCreReiGor-PRB-13}
F.~Malet, A.~Mirtschink, J.~C. Cremon, S.~M. Reimann and P.~Gori-Giorgi,
  \emph{Phys. Rev. B}, 2013, \textbf{{87}}, 115146\relax
\mciteBstWouldAddEndPuncttrue
\mciteSetBstMidEndSepPunct{\mcitedefaultmidpunct}
{\mcitedefaultendpunct}{\mcitedefaultseppunct}\relax
\EndOfBibitem
\bibitem[Mendl \emph{et~al.}(2014)Mendl, Malet, and
  Gori-Giorgi]{MenMalGor-PRB-14}
C.~B. Mendl, F.~Malet and P.~Gori-Giorgi, \emph{arXiv:1311.6011}, 2014\relax
\mciteBstWouldAddEndPuncttrue
\mciteSetBstMidEndSepPunct{\mcitedefaultmidpunct}
{\mcitedefaultendpunct}{\mcitedefaultseppunct}\relax
\EndOfBibitem
\bibitem[Mendl and Lin(2013)]{MenLin-PRB-13}
C.~B. Mendl and L.~Lin, \emph{Phys. Rev. B}, 2013, \textbf{87}, 125106\relax
\mciteBstWouldAddEndPuncttrue
\mciteSetBstMidEndSepPunct{\mcitedefaultmidpunct}
{\mcitedefaultendpunct}{\mcitedefaultseppunct}\relax
\EndOfBibitem
\bibitem[Buttazzo \emph{et~al.}(2012)Buttazzo, {De Pascale}, and
  Gori-Giorgi]{ButDepGor-PRA-12}
G.~Buttazzo, L.~{De Pascale} and P.~Gori-Giorgi, \emph{Phys. Rev. A}, 2012,
  \textbf{{85}}, 062502\relax
\mciteBstWouldAddEndPuncttrue
\mciteSetBstMidEndSepPunct{\mcitedefaultmidpunct}
{\mcitedefaultendpunct}{\mcitedefaultseppunct}\relax
\EndOfBibitem
\bibitem[Cotar \emph{et~al.}(2013)Cotar, Friesecke, and
  Kl\"uppelberg]{CotFriKlu-CPAM-13}
C.~Cotar, G.~Friesecke and C.~Kl\"uppelberg, \emph{Comm. Pure Appl. Math.},
  2013, \textbf{66}, 548\relax
\mciteBstWouldAddEndPuncttrue
\mciteSetBstMidEndSepPunct{\mcitedefaultmidpunct}
{\mcitedefaultendpunct}{\mcitedefaultseppunct}\relax
\EndOfBibitem
\bibitem[Friesecke \emph{et~al.}(2013)Friesecke, Mendl, Pass, Cotar, and
  Kl\"uppelberg]{FriMenPasCotKlu-JCP-13}
G.~Friesecke, C.~B. Mendl, B.~Pass, C.~Cotar and C.~Kl\"uppelberg, \emph{J.
  Chem. Phys.}, 2013, \textbf{139}, 164109\relax
\mciteBstWouldAddEndPuncttrue
\mciteSetBstMidEndSepPunct{\mcitedefaultmidpunct}
{\mcitedefaultendpunct}{\mcitedefaultseppunct}\relax
\EndOfBibitem
\bibitem[Helbig \emph{et~al.}(2011)Helbig, Fuks, Casula, Verstraete, Marques,
  Tokatly, and Rubio]{HelFukCasVerMarTokRub-PRA-11}
N.~Helbig, J.~I. Fuks, M.~Casula, M.~J. Verstraete, M.~A.~L. Marques, I.~V.
  Tokatly and A.~Rubio, \emph{Phys. Rev. A}, 2011, \textbf{83}, 032503\relax
\mciteBstWouldAddEndPuncttrue
\mciteSetBstMidEndSepPunct{\mcitedefaultmidpunct}
{\mcitedefaultendpunct}{\mcitedefaultseppunct}\relax
\EndOfBibitem
\bibitem[Wagner \emph{et~al.}(2012)Wagner, Stoudenmire, Burke, and
  White]{WagStoBurWhi-PCCP-12}
L.~O. Wagner, E.~M. Stoudenmire, K.~Burke and S.~R. White, \emph{Phys. Chem.
  Chem. Phys.}, 2012, \textbf{{14}}, 8581\relax
\mciteBstWouldAddEndPuncttrue
\mciteSetBstMidEndSepPunct{\mcitedefaultmidpunct}
{\mcitedefaultendpunct}{\mcitedefaultseppunct}\relax
\EndOfBibitem
\bibitem[Levy(1979)]{Lev-PNAS-79}
M.~Levy, \emph{Proc. Natl. Acad. Sci. U.S.A.}, 1979, \textbf{76}, 6062\relax
\mciteBstWouldAddEndPuncttrue
\mciteSetBstMidEndSepPunct{\mcitedefaultmidpunct}
{\mcitedefaultendpunct}{\mcitedefaultseppunct}\relax
\EndOfBibitem
\bibitem[Lieb(1983)]{Lie-IJQC-83}
E.~H. Lieb, \emph{Int. J. Quantum. Chem.}, 1983, \textbf{{24}}, 24\relax
\mciteBstWouldAddEndPuncttrue
\mciteSetBstMidEndSepPunct{\mcitedefaultmidpunct}
{\mcitedefaultendpunct}{\mcitedefaultseppunct}\relax
\EndOfBibitem
\bibitem[Seidl \emph{et~al.}(2007)Seidl, Gori-Giorgi, and
  Savin]{SeiGorSav-PRA-07}
M.~Seidl, P.~Gori-Giorgi and A.~Savin, \emph{Phys. Rev. A}, 2007,
  \textbf{{75}}, 042511\relax
\mciteBstWouldAddEndPuncttrue
\mciteSetBstMidEndSepPunct{\mcitedefaultmidpunct}
{\mcitedefaultendpunct}{\mcitedefaultseppunct}\relax
\EndOfBibitem
\bibitem[Gori-Giorgi \emph{et~al.}(2009)Gori-Giorgi, Vignale, and
  Seidl]{GorVigSei-JCTC-09}
P.~Gori-Giorgi, G.~Vignale and M.~Seidl, \emph{J. Chem. Theory Comput.}, 2009,
  \textbf{{5}}, 743\relax
\mciteBstWouldAddEndPuncttrue
\mciteSetBstMidEndSepPunct{\mcitedefaultmidpunct}
{\mcitedefaultendpunct}{\mcitedefaultseppunct}\relax
\EndOfBibitem
\bibitem[Mirtschink \emph{et~al.}(2012)Mirtschink, Seidl, and
  Gori-Giorgi]{MirSeiGor-JCTC-12}
A.~Mirtschink, M.~Seidl and P.~Gori-Giorgi, \emph{J. Chem. Theory Comput.},
  2012, \textbf{8}, 3097\relax
\mciteBstWouldAddEndPuncttrue
\mciteSetBstMidEndSepPunct{\mcitedefaultmidpunct}
{\mcitedefaultendpunct}{\mcitedefaultseppunct}\relax
\EndOfBibitem
\bibitem[Langreth and Perdew(1975)]{LanPer-SSC-75}
D.~C. Langreth and J.~P. Perdew, \emph{Solid State Commun.}, 1975,
  \textbf{{17}}, 1425\relax
\mciteBstWouldAddEndPuncttrue
\mciteSetBstMidEndSepPunct{\mcitedefaultmidpunct}
{\mcitedefaultendpunct}{\mcitedefaultseppunct}\relax
\EndOfBibitem
\bibitem[Seidl \emph{et~al.}(1999)Seidl, Perdew, and Levy]{SeiPerLev-PRA-99}
M.~Seidl, J.~P. Perdew and M.~Levy, \emph{Phys. Rev. A}, 1999, \textbf{{59}},
  51\relax
\mciteBstWouldAddEndPuncttrue
\mciteSetBstMidEndSepPunct{\mcitedefaultmidpunct}
{\mcitedefaultendpunct}{\mcitedefaultseppunct}\relax
\EndOfBibitem
\bibitem[Seidl \emph{et~al.}(2000)Seidl, Perdew, and Kurth]{SeiPerKur-PRA-00}
M.~Seidl, J.~P. Perdew and S.~Kurth, \emph{Phys. Rev. A}, 2000, \textbf{{62}},
  012502\relax
\mciteBstWouldAddEndPuncttrue
\mciteSetBstMidEndSepPunct{\mcitedefaultmidpunct}
{\mcitedefaultendpunct}{\mcitedefaultseppunct}\relax
\EndOfBibitem
\bibitem[Liu and Burke(2009)]{LiuBur-JCP-09}
Z.~F. Liu and K.~Burke, \emph{J. Chem. Phys.}, 2009, \textbf{{131}},
  124124\relax
\mciteBstWouldAddEndPuncttrue
\mciteSetBstMidEndSepPunct{\mcitedefaultmidpunct}
{\mcitedefaultendpunct}{\mcitedefaultseppunct}\relax
\EndOfBibitem
\bibitem[Seidl \emph{et~al.}(2000)Seidl, Perdew, and Kurth]{SeiPerKur-PRL-00}
M.~Seidl, J.~P. Perdew and S.~Kurth, \emph{Phys. Rev. Lett.}, 2000,
  \textbf{{84}}, 5070\relax
\mciteBstWouldAddEndPuncttrue
\mciteSetBstMidEndSepPunct{\mcitedefaultmidpunct}
{\mcitedefaultendpunct}{\mcitedefaultseppunct}\relax
\EndOfBibitem
\bibitem[R\"as\"anen \emph{et~al.}(2011)R\"as\"anen, Seidl, and
  Gori-Giorgi]{RasSeiGor-PRB-11}
E.~R\"as\"anen, M.~Seidl and P.~Gori-Giorgi, \emph{Phys. Rev. B}, 2011,
  \textbf{{83}}, 195111\relax
\mciteBstWouldAddEndPuncttrue
\mciteSetBstMidEndSepPunct{\mcitedefaultmidpunct}
{\mcitedefaultendpunct}{\mcitedefaultseppunct}\relax
\EndOfBibitem
\bibitem[Buijse \emph{et~al.}(1989)Buijse, Baerends, and
  Snijders]{BuiBaeSni-PRA-89}
M.~A. Buijse, E.~J. Baerends and J.~G. Snijders, \emph{Phys. Rev. A}, 1989,
  \textbf{{40}}, 4190\relax
\mciteBstWouldAddEndPuncttrue
\mciteSetBstMidEndSepPunct{\mcitedefaultmidpunct}
{\mcitedefaultendpunct}{\mcitedefaultseppunct}\relax
\EndOfBibitem
\bibitem[Helbig \emph{et~al.}(2009)Helbig, Tokatly, and
  Rubio]{HelTokRub-JCP-09}
N.~Helbig, I.~V. Tokatly and A.~Rubio, \emph{J. Chem. Phys.}, 2009,
  \textbf{131}, 224105\relax
\mciteBstWouldAddEndPuncttrue
\mciteSetBstMidEndSepPunct{\mcitedefaultmidpunct}
{\mcitedefaultendpunct}{\mcitedefaultseppunct}\relax
\EndOfBibitem
\bibitem[Perdew and Schmidt(2001)]{PerSmi-INC-01}
J.~P. Perdew and K.~Schmidt, \emph{Density Functional Theory and Its
  Application to Materials}, AIP Press, Melville, New York, 2001\relax
\mciteBstWouldAddEndPuncttrue
\mciteSetBstMidEndSepPunct{\mcitedefaultmidpunct}
{\mcitedefaultendpunct}{\mcitedefaultseppunct}\relax
\EndOfBibitem
\bibitem[Perdew \emph{et~al.}(2005)Perdew, Ruzsinszky, Tao, Staroverov,
  Scuseria, and Csonka]{PerRuzTaoStaScuCso-JCP-05}
J.~P. Perdew, A.~Ruzsinszky, J.~Tao, V.~N. Staroverov, G.~E. Scuseria and G.~I.
  Csonka, \emph{J. Chem. Phys.}, 2005, \textbf{123}, 062201\relax
\mciteBstWouldAddEndPuncttrue
\mciteSetBstMidEndSepPunct{\mcitedefaultmidpunct}
{\mcitedefaultendpunct}{\mcitedefaultseppunct}\relax
\EndOfBibitem
\bibitem[Levy and Perdew(1985)]{LevPer-PRA-85}
M.~Levy and J.~P. Perdew, \emph{Phys. Rev. A}, 1985, \textbf{32}, 2010\relax
\mciteBstWouldAddEndPuncttrue
\mciteSetBstMidEndSepPunct{\mcitedefaultmidpunct}
{\mcitedefaultendpunct}{\mcitedefaultseppunct}\relax
\EndOfBibitem
\bibitem[Perdew \emph{et~al.}(2004)Perdew, Tao, Staroverov, and
  Scuseria]{PerTaoStaScu-JCP-04}
J.~P. Perdew, J.~Tao, V.~N. Staroverov and G.~E. Scuseria, \emph{J. Chem.
  Phys.}, 2004, \textbf{120}, 6898--6911\relax
\mciteBstWouldAddEndPuncttrue
\mciteSetBstMidEndSepPunct{\mcitedefaultmidpunct}
{\mcitedefaultendpunct}{\mcitedefaultseppunct}\relax
\EndOfBibitem
\bibitem[Mirtschink \emph{et~al.}(2014)Mirtschink, Umrigar, {Morgan III}, and
  Gori-Giorgi]{MirUmrMorGor-JCP-14}
A.~Mirtschink, C.~J. Umrigar, J.~D. {Morgan III} and P.~Gori-Giorgi,
  \emph{submitted to J. Chem. Phys.}, 2014\relax
\mciteBstWouldAddEndPuncttrue
\mciteSetBstMidEndSepPunct{\mcitedefaultmidpunct}
{\mcitedefaultendpunct}{\mcitedefaultseppunct}\relax
\EndOfBibitem
\end{mcitethebibliography}
\end{document}